# Voltage-based magnetization switching and reading in magnetoelectric spin-orbit nanodevices

Diogo C. Vaz[1,♦], Chia-Ching Lin[2], John Plombon[2], Won Young Choi[1,*], Inge Groen[1], Isabel C. Arango[1], Andrey Chuvilin[1,3], Luis E. Hueso[1,3], Dmitri E. Nikonov[2], Hai Li[2], Punyashloka Debashis[2], Scott B. Clendenning[2], Tanay A. Gosavi[2], Yen-Lin Huang[4], Bhagwati Prasad[4,5], Ramamoorthy Ramesh[4], Aymeric Vecchiola[6], Manuel Bibes[6], Karim Bouzehouane[6], Stephane Fusil[6], Vincent Garcia[6], Ian A. Young[2], Fèlix Casanova[1,3,♦]

[1] CIC nanoGUNE BRTA, 20018 Donostia-San Sebastian, Basque Country, Spain
[2] Components Research, Intel Corp., Hillsboro, Oregon 97124, USA
[3] IKERBASQUE, Basque Foundation for Science, 48009 Bilbao, Basque Country, Spain
[4] Department of Physics, University of California, Berkeley, California 94720, USA
[5] Materials Engineering Department, Indian Institute of Science, Bengaluru, 560012, Karnataka, India
[6] Laboratoire Albert Fert, CNRS, Thales, Université Paris-Saclay, 91767, Palaiseau, France
[*] Current affiliation: VanaM Inc., 21-1 Doshin-ro 4-gil, Yeongdeungpo-gu, Seoul, Republic of Korea

As CMOS technologies face challenges in dimensional and voltage scaling, the demand for novel logic devices has never been greater, with spin-based devices offering scaling potential, at the cost of significantly high switching energies. Alternatively, magnetoelectric materials are predicted to enable low-power magnetization control, a solution with limited device-level results. Here, we demonstrate voltage-based magnetization switching and reading in nanodevices at room temperature, enabled by exchange coupling between multiferroic $BiFeO_3$ and ferromagnetic CoFe, for writing, and spin-to-charge current conversion between CoFe and Pt, for reading. We show that upon the electrical switching of the $BiFeO_3$, the magnetization of the CoFe can be reversed, giving rise to different voltage outputs. Through additional microscopy techniques, magnetization reversal is linked with the polarization state and antiferromagnetic cycloid propagation direction in the $BiFeO_3$. This study constitutes the building block for magnetoelectric spin-orbit logic, opening a new avenue for low-power beyond-CMOS technologies.

[♦] Correspondence to: d.vaz@nanogune.eu, f.casanova@nanogune.eu

**Introduction**

After 50 years of continuous transistor size downscaling and increased performance[1], future iterations of logic circuits will require beyond-CMOS alternatives[2] that explore new physical effects through non-conventional materials. While Moore's law is still sustained by increasingly complex transistor designs and lithography advances[3], the last decade saw a clear breakdown of Dennard's scaling, where smaller transistors no longer mean smaller operational voltages, compromising the energy efficiency and performance of future chips. In recent years, a flurry of new logic devices has emerged, driven by the usage of alternative state variables, such as spin, polarization, strain, and orbital[4]. Among these options, spin-based solutions have shown tremendous promise and applicability[5]. Owing to their non-volatile nature, effects like spin-transfer torque (STT) and spin-orbit torque (SOT) brought major improvements in stand-by power, as well as in terms of endurance, writing speed, and compatibility with back-end of line (BEOL) fabrication processes[6,7]. Yet, controlling magnetization states using these methods still requires rather large currents, preventing their usage as a realistic non-volatile logic solution. Alternatively, voltage-based methods gained some traction in recent years[8], mainly pushed by voltage-controlled magnetic anisotropy (VCMA)[9], where voltage-induced dynamic switching of magnetization has been reported[10]. While field-free VCMA writing has been recently shown[11], further work is required to improve the VCMA coefficient, in order to bring this technology closer to product applications.

A pathway for field-free voltage-based switching of magnetism has been proposed using magnetoelectric multiferroics[12]. Among several possible combinations, the coexistence of ferroelectricity and ferromagnetism is expected to allow the control of magnetization through switching of the ferroelectric polarization with an electric field. In this category, $BiFeO_3$ has been the most studied material, exhibiting a tight coupling between antiferromagnetic (AF) and ferroelectric (FE) orders at room temperature. One of the most notable results towards multiferroic-based devices was the demonstration of magnetization reversal by 180º in a CoFe element, exchange coupled with $BiFeO_3$, upon application of an electric field[13]. The result was interpreted considering weak ferromagnetism arising from canting of the $Fe^{3+}$ magnetic moments in $BiFeO_3$ [[14]], which can couple to the magnetization of the CoFe. Upon a two-step switching of the polarization and canted magnetization in $BiFeO_3$, the magnetization of the CoFe is expected to follow this motion and reverse[15].

Since then, the road to multiferroic-based devices has been long and tortuous, with sparse results reported[16]. Yet, it is expected that such devices can bring magnetization writing energies down to the aJ range[17], an improvement of several orders of magnitude when compared with state-of-the-art current-based devices. This driving force led to the recent proposal of magnetoelectric spin-orbit (MESO) logic[17], suggesting a spin-based nanodevice adjacent to a multiferroic, where the magnetization is switched solely with a voltage and is electrically read using spin-to-charge current conversion (SCC) phenomena.

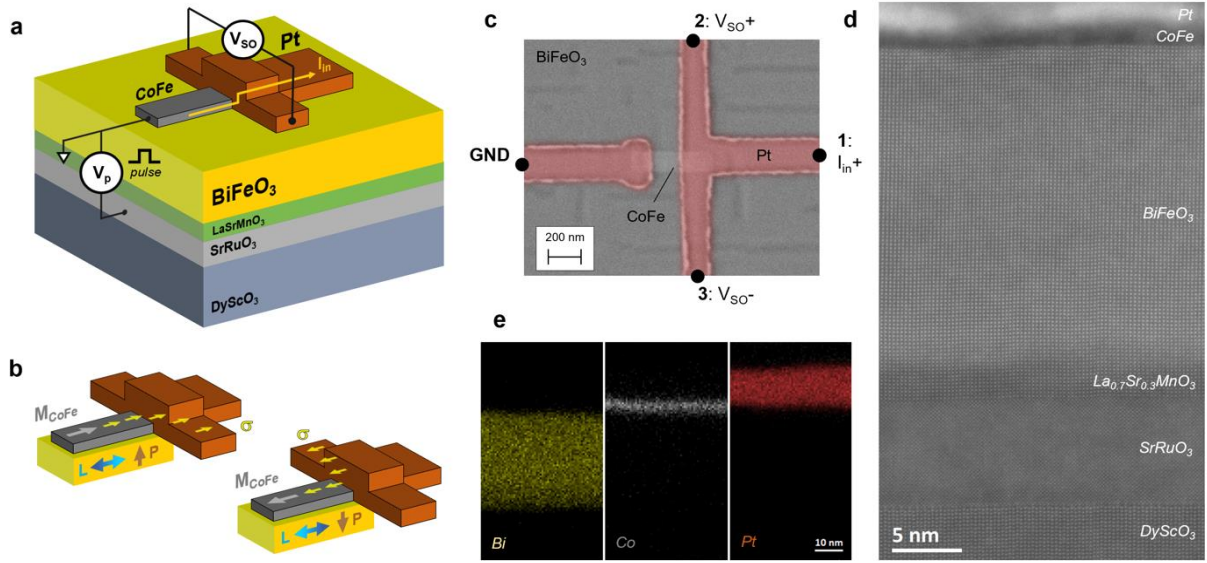

**Figure 1 – MESO nanodevice and material characterization**. **a**, MESO device configuration composed of a DyScO$_3$ substrate, La$_{0.7}$Sr$_{0.3}$MnO$_3$/SrRuO$_3$ bottom electrodes, multiferroic BiFeO$_3$, magnetic CoFe element and a high spin-orbit material Pt. The logic state variable is given by the magnetization direction in CoFe. **b**, Writing is achieved by applying voltage pulses V$_p$ between the CoFe and the bottom electrode, switching the polarization P and AF order L of BiFeO$_3$, which reverses the magnetization M$_{CoFe}$ of CoFe. Reading is achieved through SCC phenomena, where a spin-polarized current I$_{in}$ is injected into Pt, leading to different transverse output voltages V$_{SO}$, depending on the initial orientation of the injected spins $\sigma$. **c**, SEM top-view image of the fabricated nanodevice. I$_{in}$ is applied between lead 1 and ground GND, and V$_{SO}$ is detected between leads 2 and 3. **d**, TEM cross-sectional image at the Pt/CoFe junction region on the fabricated nanodevice. **e**, EDX elemental maps of Bi (from the BiFeO$_3$ layer), Co (from the CoFe layer) and Pt at the Pt/CoFe junction region.

In this article, we demonstrate the experimental implementation of such a device. We fabricate SCC nanodevices on BiFeO$_3$ and analyze the reversibility of the magnetization of CoFe using a combination of piezoresponse (PFM) and magnetic force microscopy (MFM), where the polarization state of the BiFeO$_3$ and the magnetization of CoFe are imaged upon switching. We then correlate this with all-electrical SCC experiments where voltage pulses are applied to switch the BiFeO$_3$ and reverse the magnetization of CoFe (writing), and SCC outputs different voltages depending on the magnetization direction (reading). Lastly, we investigate the magnetic textures at the surface of BiFeO$_3$ using scanning nitrogen-vacancy (NV) magnetometry, where the coupling between CoFe and BiFeO$_3$ is linked with the AF cycloid propagation direction.

**Results**

In Fig. 1a, we show a sketch of the fabricated MESO nanodevice. The MESO concept can be described as an assembly of a magnetoelectric (ME) module used for writing, and a spin-orbit (SO) module used for reading[18]. The ME module comprises the multiferroic and an adjacent ferromagnet, here BiFeO$_3$ and CoFe, respectively. Voltages pulses (V$_p$) are applied between a metallic La$_{0.7}$Sr$_{0.3}$MnO$_3$/SrRuO$_3$ bottom electrode and the CoFe, so that the polarization (P) and the AF order (L) in the BiFeO$_3$ can be switched, as exemplified in Fig. 1b. Here, the magnetization direction of CoFe (M$_{CoFe}$) is also reversed, following the reversal of P and L, due to exchange coupling at the CoFe/BiFeO$_3$ interface. The SO module is based on a T-shaped nanostructured device composed of CoFe and the spin-orbit material Pt, following a recent study on SCC for magnetic logic readout[19]. A spin-polarized current (I$_{in}$) is electrically driven from CoFe to Pt, where, at the Pt/CoFe junction, the spins are converted into a charge current through the inverse spin Hall effect (ISHE) and picked up as a transverse voltage V$_{SO}$. Depending on the magnetization direction, spins σ are deflected either to the right or left, as shown in Fig. 1b, enabling a fully electrical method of magnetization state readout, that generates an electromotive force that can drive another circuit element. The device is based on a Pt(10 nm)/CoFe(2.5 nm)/BiFeO$_3$(30 nm)/La$_{0.7}$Sr$_{0.3}$MnO$_3$(4 nm)/SrRuO$_3$(10 nm) stack grown on a DyScO$_3$ (110) substrate, using a combination of pulsed laser deposition and sputtering (see details in Methods). The fabrication of the device comprises positive nanolithography processes using e-beam lithography, Ar-ion milling, and sputtering, used to define both the CoFe wire (500 nm x 150 nm in lateral size), the Pt T-shaped electrode, and contacts. The device is capped with SiO$_2$(5 nm) to prevent oxidation of the CoFe. Details of the fabrication process flow can be found in Supplementary Information Note 1. A scanning electron microscopy (SEM) top image of the integrated MESO nanodevice is shown in Fig. 1c, and a cross-sectional image of the Pt/CoFe junction area, taken by transmission electron microscopy (TEM) after device fabrication, is shown in Fig. 1d. We observe highly ordered epitaxial growth of the oxide heterostructure, as well as clean and sharp interfaces between BiFeO$_3$, CoFe, and Pt. From the energy dispersive X-ray spectroscopy (EDX) maps shown in Fig. 1e, we observe minimal interdiffusion between the three layers.

We start by investigating the magnetization orientation of CoFe upon polarization reversal of the BiFeO$_3$ using a combination of PFM and MFM. In Fig. 2a, we observe that the polarization of the bare BiFeO$_3$ can be poled up (dark area) and down (bright area) with positive (2 V) and negative (−2 V) voltages, respectively. Then, a CoFe/Pt disk 5 μm in diameter (patterned similarly to the MESO devices) was used to measure the current and polarization vs. voltage loops, as shown in Fig. 2b and 2c, respectively. These capacitors based on a 30-nm-thick BiFeO$_3$ show large saturation polarization (close to the bulk value), with low leakage, as well as low switching voltages. Indeed, we observe switching voltages of −0.5 V and 1.5 V underneath the metallic disk. A relatively large imprint, normally observed in thin ferroelectric films due to top and bottom contact electrostatic asymmetries, is still present, even though largely improved by the $La_{0.7}Sr_{0.3}MnO_3$ buffer layer[20]. As illustrated in Fig. 2d, disks 300 nm in diameter were then used to evaluate both the out-of-plane ($P_{out}$) and in-plane ($P_{in}$) polarization direction in the $BiFeO_3$ underneath the disk, as well as the direction of $M_{CoFe}$, labeled as V-PFM, L-PFM, and MFM in Fig. 2e, respectively (setup details in Methods).

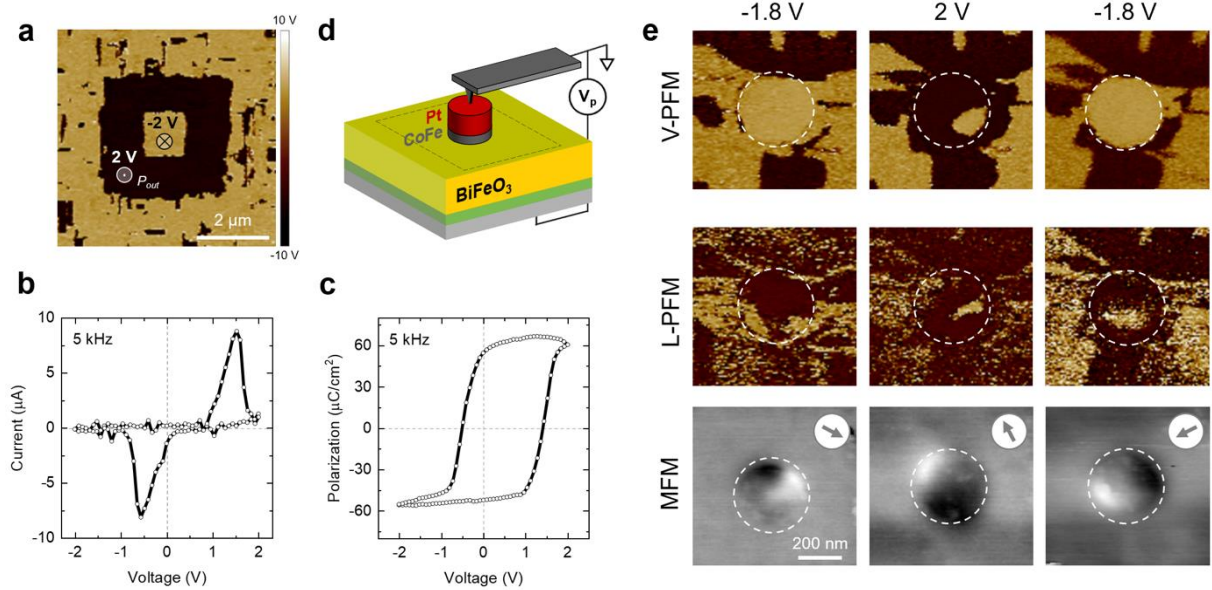

**Figure 2 – PFM and MFM characterizations. a**, Out-of-plane polarization $P_{out}$ after a box-in-box switching experiment on the bare $BiFeO_3$ surface. Dark and bright areas correspond to polarization poled up and down, respectively. **b**, Current and **c**, polarization vs. voltage loops on Pt/CoFe disks 5 μm in diameter over $BiFeO_3/SrRuO_3/DyScO_3$, collected with a frequency of 5 kHz. **d**, Sketch of the PFM and MFM experiments. Dashed line corresponds to the area scanned with PFM and MFM. **e**, Out-of-plane (V) and in-plane (L) PFM phase images after applying voltage pulses of −1.8 V, 2 V, and −1.8 V to a Pt/CoFe disk 300 nm in diameter, showing the FE domains in $BiFeO_3$ underneath the disk. Corresponding MFM images showing the magnetization direction of the CoFe after each pulse, represented by the grey arrows.

From V-PFM data, we observe that $P_{out}$ can be reversed between down and up states, after applying voltage pulses of −1.8 V and 2 V, respectively. Small unswitched patches were occasionally observed, either caused by incomplete switching after $V_p$ is applied or by relaxation back to the P down state after the voltage pulse is applied (V = 0 V) due to the $BiFeO_3$ imprint. While $P_{out}$ reverses consistently, we observe from L-PFM data that $P_{in}$ remains split into randomly distributed FE domains for the three voltage pulses applied. For 9 different devices probed in the same sample, $P_{out}$ is always switched, while $P_{in}$ exhibits mostly slight changes in the FE domain structure, suggesting a combination of local 71º/109º/180º switch of the polarization[13]. From MFM, we observe that after poling the $BiFeO_3$ down with $V_p$=−1.8 V, $M_{CoFe}$ points diagonally to the bottom right. Poling the $BiFeO_3$ up with $V_p$=2 V reverses the magnetization of CoFe by nearly 180°. Poling the polarization back down with $V_p=-1.8$ V drives a rotation of $M_{CoFe}$ by ~90°. Out of the 9 devices tested, the magnetization could always be switched in 3 of them (33.5%), could be partially/randomly switched in 4 (44.5%), and could never be switched in 2 (22%). Out of 24 out-of-plane polarization switching events, we observed that the magnetization switched 13 times (54%) and did not switch 11 times (46%). Extended data and additional switching experiments can be found in Supplementary Information Note 2.

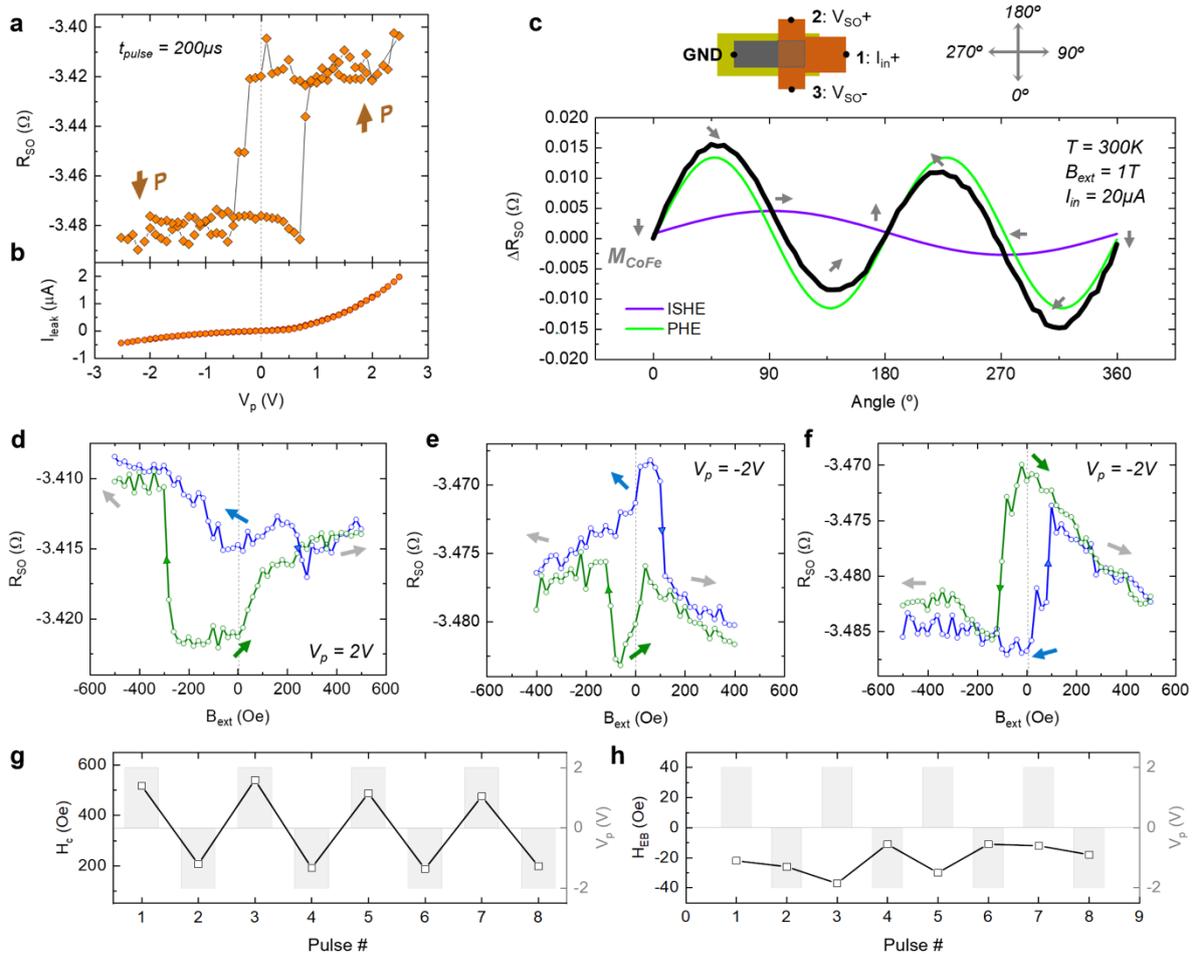

**Figure 3 – Electrical characterization of MESO nanodevices. a**, Baseline of the output resistance $R_{SO}$ and **b**, leakage current $I_{leak}$ as a function of the voltage pulse $V_p$ applied between the Pt/CoFe nanodevice and the back of the $BiFeO_3$. Two resistance states are visible depending on the polarization P. Resistances are collected 1 s after the pulse is applied. **c**, $\Delta R_{SO}=R_{SO}-R^0_{SO}$ as a function of the angle of an in-plane external magnetic field $B_{ext}=1$ T (black curve), after $V_p=-2$ V, where $R^0_{SO}$ represents the resistance measured at 0°. The data is decomposed in an inverse spin Hall effect (ISHE, in violet) and a planar Hall effect (PHE, in green) component. Grey arrows represent the magnetization of CoFe as seen from the top (see top-view sketch above). **d**, $R_{SO}$ as a function of $B_{ext}$ applied along the long axis of CoFe, after $V_p=2$ V, **e**, $V_p=-2$ V (inhomogeneous coupling) and **f**, $V_p=-2$ V (fully reversed). The blue and green curves correspond to a $B_{ext}$ sweep from −500 Oe to 500 Oe and back, respectively. Arrows represent $M_{CoFe}$ as seen from the top. **g**, Coercivity $H_c$ and **h**, exchange bias $H_{EB}$ of the CoFe element as function of different voltage pulses alternating between $V_p=2$ V and $V_p=-2$ V (grey bars).

Given these results, we conclude that the reversal of $M_{CoFe}$ is still possible even though there is a lack of control of $P_{in}$, which should be intimately related to the in-plane component of the AF order and canted magnetization in $BiFeO_3$. Moreover, $P_{out}$ switching seems to be driving the reversal/rotation of $M_{CoFe}$, but does not always guarantee it, indicating that the magnetic configuration in a uniformly out-of-plane polarized region may be more complex, a hypothesis investigated further ahead in this article. Regardless, these results reveal that the magnetization can be manipulated in nanoscale magnets interfaced with $BiFeO_3$ using only a voltage pulse and without external magnetic fields, experimentally demonstrating the MESO writing capabilities.

We now move to the electrical experiments on the MESO nanodevices. First, we investigate the switching ability of the $BiFeO_3$ by applying voltage pulses with a duration of 200 μs between the CoFe element and the bottom of the $BiFeO_3$. After each pulse, we perform SCC experiments on the nanodevice, as illustrated in Fig. 1c, by applying $I_{in}$=20 μA and measuring the output voltage $V_{SO}$, hereinafter shown as a resistance $R_{SO}=V_{SO}/I_{in}$. As shown in Fig. 3a, upon switching the $BiFeO_3$, the baseline of the SCC signal acquires two stable states, −3.48 Ω and −3.42 Ω. This baseline resistance reflects the slight misalignment of the CoFe element with respect to the Pt T-shaped electrode, giving rise to either a positive or negative transverse voltage[19]. The shift in baseline resistance can be explained by slight modulation of the resistivity of CoFe, either due to a static field effect from the remanent polarization in the $BiFeO_3$ (Ref.[21]), or strain induced by different ferroelastic domains. While this resistance vs. voltage loop does not give quantitative information about the polarization, we observe that the $BiFeO_3$ directly underneath the nanodevices switches at −350 mV and 750 mV, in fair agreement with the results from PFM. As shown in Fig. 3b, the leakage current measured through the $BiFeO_3$ layer during the voltage pulses was minimized to about 0.5-1 μA (for $V_p=\pm 2$ V), largely due to the reduced fabricated area of CoFe and Pt in direct contact with $BiFeO_3$.

To fully characterize the SCC results with respect to the expected $M_{CoFe}$ orientation, we measure $\Delta R_{SO}$ as a function of a rotating in-plane external magnetic field $B_{ext}$=1 T, enough to fully saturate the CoFe element (Fig. 3c). Here, $\Delta R_{SO}$ represents the relative variation of $R_{SO}$ as the magnetization rotates, allowing a correspondence with the $R_{SO}$ changes observed in Fig. 3d-f. Grey arrows indicate the $M_{CoFe}$ orientation as seen from above. $R_{SO}$ is dominated by the ISHE at 90º and 270º (in violet) when $M_{CoFe}$ points along the easy axis of the CoFe element, following a $\sin(\theta)$, and by the planar Hall effect (PHE) at diagonal orientations (45º, 135º, 225º, and 315º) (in green), following a $\sin(2\theta)$. Similar behavior was identified in Ref.[22]. In Fig. 3 d-f, we show $R_{SO}$ as a function of $B_{ext}$ along the CoFe long (easy) axis (90º), after voltage pulses of $V_p=\pm 2$ V. Voltage pulses are applied without any external magnetic field, and the loops are taken by sweeping $B_{ext}$ from −500 Oe to 500 Oe (in blue) and back (in green). These loops only provide information on the inherent interactions between $M_{CoFe}$ and $BiFeO_3$, while the $M_{CoFe}$ direction manipulation right after different $V_p$ is investigated further ahead. After applying $V_p$=2 V, the $B_{ext}$ sweep reveals that $R_{SO}$ decreases when $B_{ext}$ approaches zero, indicating that, when mapped to the angle dependence, the magnetization tilts up (Fig. 3d). After applying $V_p$=−2 V, we observe two possible states. In the first case, $M_{CoFe}$ also tilts up around $B_{ext}$=0 (Fig. 3e). We postulate that given the occasional incompleteness of the polarization switching observed from PFM, the area underneath the CoFe element may be at times split into different domains, giving rise to inhomogeneous coupling. However, in Fig. 3f the magnetization loop is reversed, and $M_{CoFe}$ tilts down around $B_{ext}$=0. Additional data concerning the reproducibility of the two fully switched magnetization loops and their correspondence with the angle dependence can be found in Supplementary Information Notes 3 and 4. Unlike T-shaped devices fabricated on Si/SiO$_2$ substrates where at zero external magnetic field $M_{CoFe}$ either points to the right (90º) or to the left (270º) due to shape anisotropy[19], on $BiFeO_3$ the CoFe magnetization may be pulled in any direction, depending on the magnetic textures underneath the CoFe. The observed tilt of $M_{CoFe}$ in our devices suggests that the exchange energy is larger than the shape anisotropy, leading to non-trivial $M_{CoFe}$ orientations in the absence of external magnetic fields. Additionally, we show in Fig. 3g that the CoFe coercivity $H_c$, obtained by the difference between switching fields, changes deterministically between ~500 Oe and 200 Oe, as observed in previous reports of exchange coupling at CoFe/$BiFeO_3$ interfaces[23,24]. However, no evident correlation is seen between $V_p$ and the exchange bias $H_{EB}$ (Fig. 3h), obtained by the sum of the switching fields, suggesting prevalent exchange coupling with the AF order[25], rather than the weak ferromagnetism in $BiFeO_3$, that would pull the magnetization in opposite directions depending on the polarization state.

Moving towards a scenario that is closer to the full implementation of MESO logic, we now investigate the $M_{CoFe}$ orientation right after applying $V_p$=±2 V. We note that, since $M_{CoFe}$ is tilted when $B_{ext}$=0 T, the reading function of the MESO device will mostly rely on the PHE instead of the ISHE. While this may reduce the overall output reading voltage, it is still sufficient to electrically probe the magnetization direction in our experiments.

We start by initializing the magnetization direction towards the right by applying a $V_p$=−2 V and sweeping the external magnetic field from 0 to 400 Oe and back. From this state, we apply $V_p$=2 V at zero magnetic field and measure $R_{SO}$ as a function of $B_{ext}$, to see to which branch of the full loop (of Fig. 3d) this half sweep corresponds. As shown in Fig. 4a, a higher initial $R_{SO}$ is observed (in blue), corresponding to a magnetization rotation by either 90º or 180º. Out of eight attempts, this behavior was observed four times (in the same device), with the remaining attempts showing no noticeable change in $R_{SO}$ (in grey).

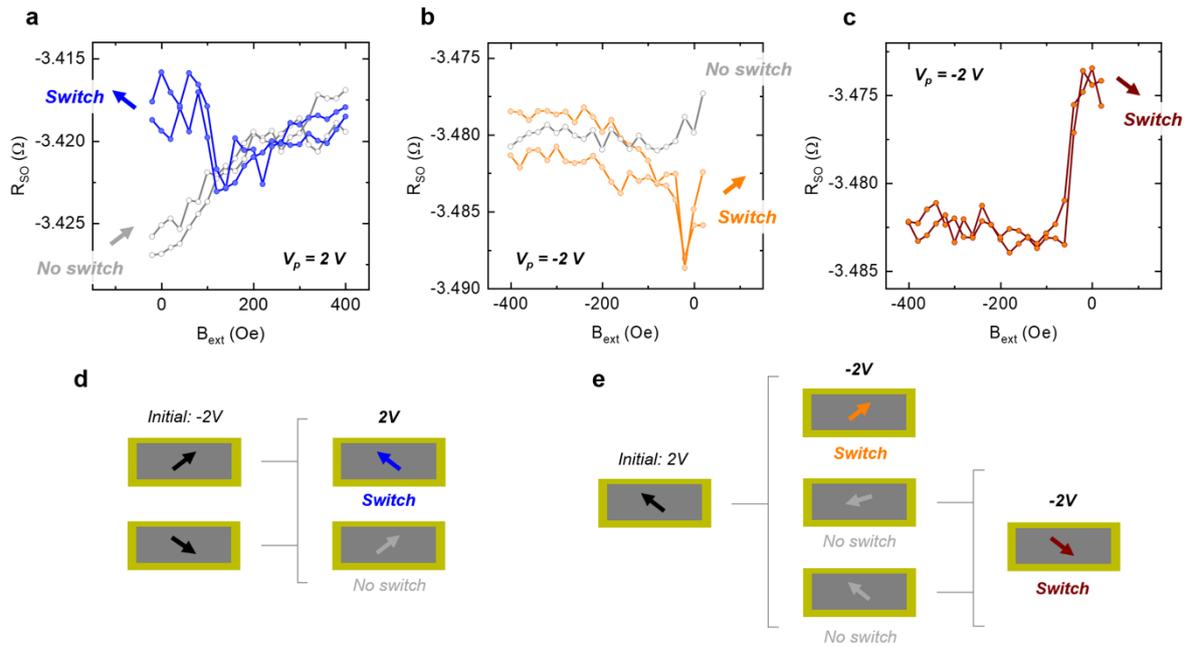

**Figure 4 – Voltage-based magnetization switching and reading in MESO nanodevices. a**, $R_{SO}$ as a function of $B_{ext}$, swept from 0 Oe to 400 Oe, after applying $V_p$=2 V. The blue curve shows a switch of $M_{CoFe}$ (arrow pointing to the top left), with a higher initial $R_{SO}$ that then reverts to the initial state with increasing $B_{ext}$. The grey curve represents a non-switch event. **b**, $R_{SO}$ as a function of $B_{ext}$, swept from 0 Oe to −400 Oe, after applying $V_p$=−2 V. The orange curve shows a switch of MCoFe (arrow pointing to the top right), with a decrease and increase of $R_{SO}$. After the first pulse, $M_{CoFe}$ is not switched (grey curve). The grey curve represents a non-switch event. **c**, After the field is swept back to 0 Oe, a second pulse $V_p$=−2 V. The red curve shows a switch of $M_{CoFe}$ (arrow pointing to the bottom right), with a higher initial $R_{SO}$ that then reverts to the initial state with increasing negative $B_{ext}$. **d, e,** Sketches of every $M_{CoFe}$ switching path after $V_p$=2 V and $V_p$=−2 V, with the initial magnetization orientation represented by the black arrows, switching events by colored arrows, and non-switching events by grey arrows.

The magnetization direction was then initialized towards the left by applying $V_p$=2 V and sweeping the magnetic field from 0 to −400 Oe and back. As shown in Fig. 4b, a first negative voltage pulse $V_p$=−2 V can lead to a magnetization sweep where $R_{SO}$ decreases and then increases (in orange), in close similarity to the lower branch of the full magnetization loop in Fig. 3e, indicating a magnetization rotation of 90º. This behavior was observed three out of nine times, with the remaining sweeps showing no special features (in grey). By bringing the magnetic field back from −400 Oe to 0 Oe, the magnetization is now realigned to the left, and a second $V_p$=−2 V is applied. This time, a higher $R_{SO}$ is measured followed by an abrupt decrease (Fig. 4c), corresponding to the higher branch of the full magnetization in Fig. 3f and to a 90º or 180º reversal of $M_{CoFe}$. This behavior was observed in eight out of nine switching attempts. All possible magnetization switching paths are illustrated in Fig. 4d and 4e, after positive and negative voltage pulses, respectively. In addition, the full switching data and statistics can be found in Supplementary Information Notes 5 and 6. All-in-all, these results demonstrate that the magnetization can be reversed and read through voltage inputs and outputs, for both positive and negative V$_p$, fulfilling the initial MESO proposal. This writing/reading functionality is further explored in Supplementary Information Note 7, where R$_{SO}$ is consecutively probed for alternating V$_p$, in a steady state, with and without a static external magnetic field applied. The uncertainty imposed by possible diagonal magnetization orientations after switching, that give rise to a mix of SHE and PHE, contributes to the difficulty in associating one single R$_{SO}$ output to a particular magnetization direction. In addition, the presence of a R$_{SO}$ baseline shift, described before, accounts for ~80 mΩ of the signal change, overlapping with the R$_{SO}$ difference between opposite magnetization orientations, expected to be in the order of 5 to 10 mΩ. Considering the baseline shift an inherent feature of the magnetic material when in direct contact with a switchable BiFeO$_3$, it is expected that future optimization of the output signal with materials beyond Pt, as discussed further ahead, should, in principle, be sufficient to surpass the magnitude of the baseline shift.

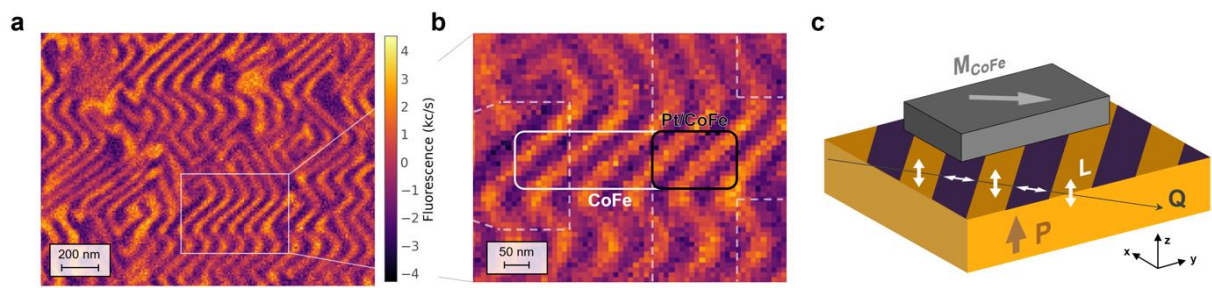

**Figure 5 – Magnetic textures and spin cycloid in BiFeO$_3$. a**, NV magnetometry images on the bare BiFeO$_3$ surface where the MESO nanodevice was fabricated. **b**, Zoomed region with a superimposed sketch (to scale) of the MESO nanodevice, revealing the possible complex magnetic behavior underneath the CoFe element. **c**, Suggested coupling mechanism between M$_{CoFe}$ and the propagation direction of the cycloid Q.

We further investigate the nature of the coupling between CoFe and BiFeO$_3$, which is expected to be responsible for the switchable M$_{CoFe}$. While the switching mechanism may be explained by coupling between M$_{CoFe}$ and the canted magnetization in BiFeO$_3$ (Ref.[13]), the spin cycloid, reported in BiFeO$_3$ thin films grown on DyScO$_3$ substrates[26,27], may complicate this interpretation. Indeed, through scanning nitrogen-vacancy (NV) magnetometry, we show in Fig. 5a that the cycloid is also present in our 30-nm-thick BiFeO$_3$, with a rotating AF order propagating in-plane in diagonal directions with a period of about ~70 nm, and changing its propagation direction (Q) by 90º in neighboring FE domains[28]. Here, the periodic variation of the magnetic stray field comes from the spin-density wave that is locked to the cycloid and perpendicular to the cycloidal plane defined by Q and P (Ref.[27,29]). As exemplified in Fig. 5b, given the dimensions of the CoFe nanostructured element in our MESO devices, represented by the white rectangle, five full rotations of the AF order within each one of the single FE domain stripes are expected to interact with M$_{CoFe}$, with two of these rotations within the Pt/CoFe junction area (black rectangle). Within this area, the canted magnetization in BiFeO$_3$ should, in principle, average to zero. Since the magnetization in the CoFe is shown to be pulled in diagonal directions in the absence of external fields, as seen from the MFM and the electrical read-out characterization, we infer that $M_{CoFe}$ may in fact couple with Q (Fig. 5c). Through this type of coupling, a rotation of Q by ~90°/180°, for a partially or fully switched $BiFeO_3$, respectively, may be responsible for the reversal/rotation of $M_{CoFe}$.

**Discussion**

We finish by discussing the reproducibility and non-deterministic aspects of our results, in light of the complex ferroelectric and magnetic textures of $BiFeO_3$. For two identical devices fabricated over different regions of the $BiFeO_3$, $M_{CoFe}$ may interact with completely different magnetic textures. Depending on this interaction, $M_{CoFe}$ may initially be pulled in different directions, so that the same voltage pulse polarity will drive different rotation/reversal paths, as observed in the PFM/MFM data in Supplementary Information Notes 2. This lack of correspondence between positive and negative voltage pulses with specific directions of $M_{CoFe}$ makes device-to-device reproducibility a real challenge. Not only that, but due to "maze-like" magnetic regions observed in Fig. 5a, some devices may not even exhibit a clear coupling with the $BiFeO_3$, a scenario further discussed in Supplementary Information Note 9. These issues may potentially be solved by better control of the ferroelectric domain structure of $BiFeO_3$ itself (or another multiferroic), ideally culminating in controlled single macroscopic domain regions with a coherent cycloid propagation[30,31] that can be effectively switched[32]. Alternatively, the overall absence of the cycloid could simplify the coupling mechanism, where $M_{CoFe}$ would couple with a uniform AF order in the multiferroic. Once these obstacles are surpassed, a systematic study would be desirable, using NV magnetometry to simultaneously probe the magnetic texture in $BiFeO_3$ and the orientation of $M_{CoFe}$, and matching the switchable cycloid[32] with the magnetization state of the magnetic element.

Besides these fundamental issues, the future implementation of MESO logic will require additional improvements on both the ME and SO modules. Unlike STT or SOT current-based solutions, the reliability of the writing on MESO devices does not improve with larger input signals. In fact, as long as the $BiFeO_3$ can be engineered to switch robustly at lower voltages, the writing energies can be progressively reduced without compromising the reliability of the writing. The key elements to consider are the coupling between the magnet and the $BiFeO_3$, together with a soft magnet that can easily "follow" the magnetic motion in the multiferroic[15], while maintaining an overall thermally stable magnetization state and FE domain structure. Further miniaturization of the magnetic and spin-orbit elements to sub-100-nm features, together with the reduction of $BiFeO_3$ thickness, switching voltages (through La doping[33,34]), leakage currents (through Mn doping[35]) and switching pulse duration (down to tens of ns[36]) are pathways to reduce the switching energies to fJ and aJ ranges. Extended discussion on pulse duration implications and reproducibility of the coupling between CoFe and $BiFeO_3$ are presented in Supplementary Information Notes 8 and 9. In terms of endurance, the bottleneck comes from the degradation of the CoFe/$BiFeO_3$ interface with voltage cycling, given the possible formation of an oxidized or intermixed interfacial magnetic layer, as well as the degradation of $BiFeO_3$ itself. Solutions such as all-oxide epitaxial structures are one possible avenue to improve this[37]. On the SO module side, SCC output voltages between opposite magnetization states need to be at least comparable with the switching voltages of $BiFeO_3$, and ideally based on the ISHE instead of PHE, to make it scalable[19]. As initially discussed in Ref. [17], MESO devices interconnected as cascaded logic gates will require the output voltage of one device to match the input switching voltage of the next one. For an input reading current of 100 µA, our SCC devices only show $\Delta V_{out}$=1 µV, while optimized Pt- and Ta-based devices reaching 30 µV and 0.35 mV, respectively, have been reported[19,38]. Besides the output voltage, an increase in the corresponding output current, resulting from more efficient spin-charge conversion phenomena, will also play an important role in faster switching. Nevertheless, additional efforts are required to reach hundreds of mV, potentially through all-electrical SCC in more exotic systems, such as topological insulators[39] and oxide heterostructures.

In conclusion, we have shown voltage-based writing and reading of magnetic states in a CoFe nanostructured element coupled with multiferroic $BiFeO_3$, representing the proof-of-principle for the MESO logic concept. Through a combination of PFM and MFM, we observe that the magnetization of CoFe can undergo 90º and 180º rotation/reversal, when the out-of-plane FE polarization of $BiFeO_3$ is switched using voltage pulses of ±2 V. Using CoFe and Pt-based T-shaped nanostructures, we electrically detected the magnetization rotation/reversal, which leads to different voltage output states depending on the direction of CoFe magnetization. The presence of a spin cycloid with a period smaller than the size of the nanostructured magnet suggests that the magnetization control is driven by coupling with the propagating AF cycloid in $BiFeO_3$. While further work is required in terms of controllability and reproducibility of the switching, specifically regarding the ferroelectric and magnetic textures in $BiFeO_3$, these results provide a key step forward towards voltage-control of magnetization in nanoscale magnets, essential for future low-power spin-based logic and memory devices.

## Methods

*Sample preparation*

The DyScO$_3$(110) substrates were purchased from MTI Corporation and were cleaned with 20 min sonication at room temperature in acetone. The DySsO$_3$ substrates were bonded onto an Inconel carrier using silver paint. The silver paint was cured on a hot plate heated to 185ºC. The SrRuO$_3$, La$_{0.3}$Sr$_{0.7}$MnO$_3$, and BiFeO$_3$ were deposited using pulsed laser deposition with a laser fluence of approximately 1.5 J/cm$^2$ at 10 Hz and oxygen pressure of 150 mTorr. The SrRuO$_3$ was deposited at 690ºC and the La$_{0.3}$Sr$_{0.7}$MnO$_3$ and BiFeO$_3$ were deposited at 650ºC to minimize the Mn diffusion. The Co and Pt were deposited by physical vapor deposition in an in-situ magnetic field of ~300 to 400 Oe. A short vacuum break after the pulsed laser deposition (less than 45 seconds) was used to place the DyScO$_3$ Inconel carrier onto the physical vapor deposition sample holder configured with permanent magnets.

*Nanodevice fabrication*

The devices were fabricated on Pt(10 nm)/CoFe(2.5 nm)/BiFeO$_3$(30 nm)/ La$_{0.7}$Sr$_{0.3}$MnO$_3$(4 nm)/SrRuO$_3$(10 nm)/DyScO$_3$(110) samples (described above) with a multiple-step e-beam lithography, metal and oxide sputtering deposition, Ar-ion milling and lift-off process. Milling of the initial CoFe/Pt is performed with the ion gun at 10º with respect to the sample surface normal, an Ar flow of 15 s.c.c.m., an acceleration voltage of 50 V, a beam current of 50 mA and a beam voltage of 300 V. Side wall milling of nanostructures after lift-off is performed in the same conditions, with an angle of the ion gun at 80º. Control of the milling rates is achieved through real time end-point mass spectrometer and resistivities of milled films. Pt T-shaped nanostructures are fabricated using a positive PMMA 950A2 e-beam resist and deposited by magnetron sputtering with a rate of 1.25 Å.s$^{-1}$, 80 W of power, 1.0×10$^{-8}$ mtorr of base pressure, 3 mtorr of Ar pressure. Isolation layer for wire bonder contact pads of Al$_2$O$_3$ is fabricated with a double-layer PMMA 495A4 + PMMA 950A2 resists and deposited with RF magnetron sputtering with a rate of 0.2 Å.s$^{-1}$, 300 W of power, 1.0×10$^{-8}$ mtorr of base pressure, 3 mtorr of Ar pressure. All lift-offs were performed using acetone.

*Transmission electron microscopy and EDX*

STEM and EDX studies were performed on Titan 60-300 Electron Microscope (FEI, Netherlands) at 300 kV accelerating voltage. The microscope was equipped by x-FEG, gun monochromator, retractable RTEM EDX detector (EDAX, USA) and HAADF detector. STEM images were acquired at nominal spot size 9, 10 mrad convergence angle, and −50 V relative gun lens potential. EDX mapping was done at a nominal spot size of 6 and −15 V gun lens potential to provide a sufficient count rate. The cross-sections of the devices were prepared by a standard FIB lamellae fabrication technique: a protective Pt layer was deposited first by e-beam followed by ion-beam deposition, lamellae of ~2 μm thickness were undercut and transferred onto a copper half-grid, thinned there to ~200 nm by 30 keV Ga$^+$ beam, and finally polished to ~20 nm at 5 keV.

*Piezoresponse force microscopy*

PFM experiments were conducted with an atomic force microscope (Nanoscope V multimode, Bruker). Two external lock-in detectors (SR830, Stanford Research) were used to simultaneously acquire vertical and lateral piezoresponses. An external source (DS360, Stanford Research) was used to excite the La$_{0.7}$Sr$_{0.3}$MnO$_3$/SrRuO$_3$ bottom electrode (ac 0.6 V peak-to-peak at 35 kHz) while the conducting Pt-coated tip was grounded. Pt-coated tips (Budget Sensors) with 40 Nm$^{-1}$ cantilevers were chosen for these images. For the experiments on the bare BiFeO$_3$ surface (Fig. 2a), the same source was used to write domains with a dc voltage while scanning. For the experiments on devices (Fig. 2c), write voltage pulses (1 sec) were applied while the tip was in contact with Pt/CoFe top electrode but not scanning.

*Magnetic force microscopy*

The MFM observation of the Pt/CoFe nanostructures has been performed in a setup under low pressure, of the order of P = $10^{-6}$ mbar. Images were obtained at room temperature using magnetic tips in a double-pass tapping–lift mode, detecting the phase shift of the second pass after a topographic measurement and thus probing the magnetic field gradient along the vertical direction. Tips were fabricated in our laboratory by depositing a magnetic coating on commercial silicon tips with magnetic sputtering, whose thicknesses were in the range 3–23 nm for CoFeB, which we selected for their particularly low degree of perturbation on the magnetic configurations under observation and improved signal-to-noise ratio, with quality factor Q = 1,500 and spring constant k = 0.4 N m$^{-1}$.

*Electrical characterization*

Transport measurements are performed in a Physical Property Measurement System from Quantum Design, using a "d.c. reversal" technique with a Keithley 2182 nanovoltmeter and a 6221 current source at 300 K. The input current $I_{in}$ for the measurements is 20 µA. Gate voltage pulses are applied with a Keithley 2636B Sourcemeter, with a pulse duration of 200 µs. Samples are mounted in a rotatable sample stage and the external magnetic field $B_{ext}$ is applied with a superconducting solenoid magnet. Devices are contacted using a wire bonder, with Au wire heated at 70 ºC and a force of 20 cN.

*Scanning NV magnetometry*

The antiferromagnetic spin textures of $BiFeO_3$ are imaged using a commercial scanning N-V magnetometer (ProteusQ™, Qnami AG) operated under ambient conditions. In our setup, the scanning tip is a commercial all-diamond probe with a single N-V defect at its apex integrated on a quartz tuning fork (Quantilever™ MX, Qnami AG). The diamond tip is integrated into a tuning-fork-based atomic force microscope combined with a confocal microscope optimized for single N-V defect spectroscopy.

**Data availability**

All data are available in the main text and the Supplementary Information. Additional data related to the findings in this study can be requested from the authors.

## Acknowledgments

We acknowledge C. Rufo, R. Llopis, and R. Gay for technical assistance with the sample fabrication and electrical characterization. This work is supported by Intel Corporation through the Semiconductor Research Corporation under MSR-INTEL TASK 2017-IN-2744 and the 'FEINMAN' Intel Science Technology Center, by the Spanish MCIN/AEI/10.13039/501100011033 and by ERDF A way of making Europe (Project No. PID2021-122511OB-I00 and Maria de Maeztu Units of Excellence Programme No. CEX2020-001038-M). D. C. Vaz acknowledges support from the European Commission for a Marie Sklodowska-Curie individual fellowship (Grant No. 892983-SPECTER). W. Y. Choi acknowledges postdoctoral fellowship support from 'Juan de la Cierva Formación' program by the Spanish MCIN/AEI (grant No. FJC2018-038580-I). The team at Laboratoire Albert Fert acknowledges the support from the French Agence Nationale de la Recherche (ANR) through the project TATOO (ANR-21-CE09-0033), the European Union's Horizon 2020 research and innovation program under the Grant Agreements No. 964931 (TSAR), a public grant overseen by the ANR as part of the "Investissements d'Avenir" program (Labex NanoSaclay, reference: ANR-10-LABX-0035), and the Sesame Ile de France IMAGeSPIN project (No. EX039175).

## Author Contributions Statement

F.C. and I.A.Y. proposed and supervised the study with the help of L.E.H., D.C.V. and C.-C.L. J.P. performed the growth and deposition of materials, with inputs from Y.-L.H., B.P. and R.R. D.C.V., W.Y.C., I. G. and I.C.A. designed and fabricated the nanodevices, with inputs from B.P., C.-C.L., D.E.N., H.L., P.D., S.B.C., T.A.G., I. A.Y., and F.C. A.C. performed the EDX and TEM analysis. A.V., K.B., S.F., and V.G. performed the PFM, MFM and NV magnetometry experiments and analyzed the data with inputs from M.B. D.C.V. performed the electrical characterization and analyzed the data with the help of all authors. D.C.V. wrote the manuscript with inputs from all authors. All authors discussed the results and contributed to their interpretation.

## Competing Interests Statement

The authors declare no competing interests.

# SUPPLEMENTARY INFORMATION

**Supplementary Note 1: Fabrication of MESO device.**

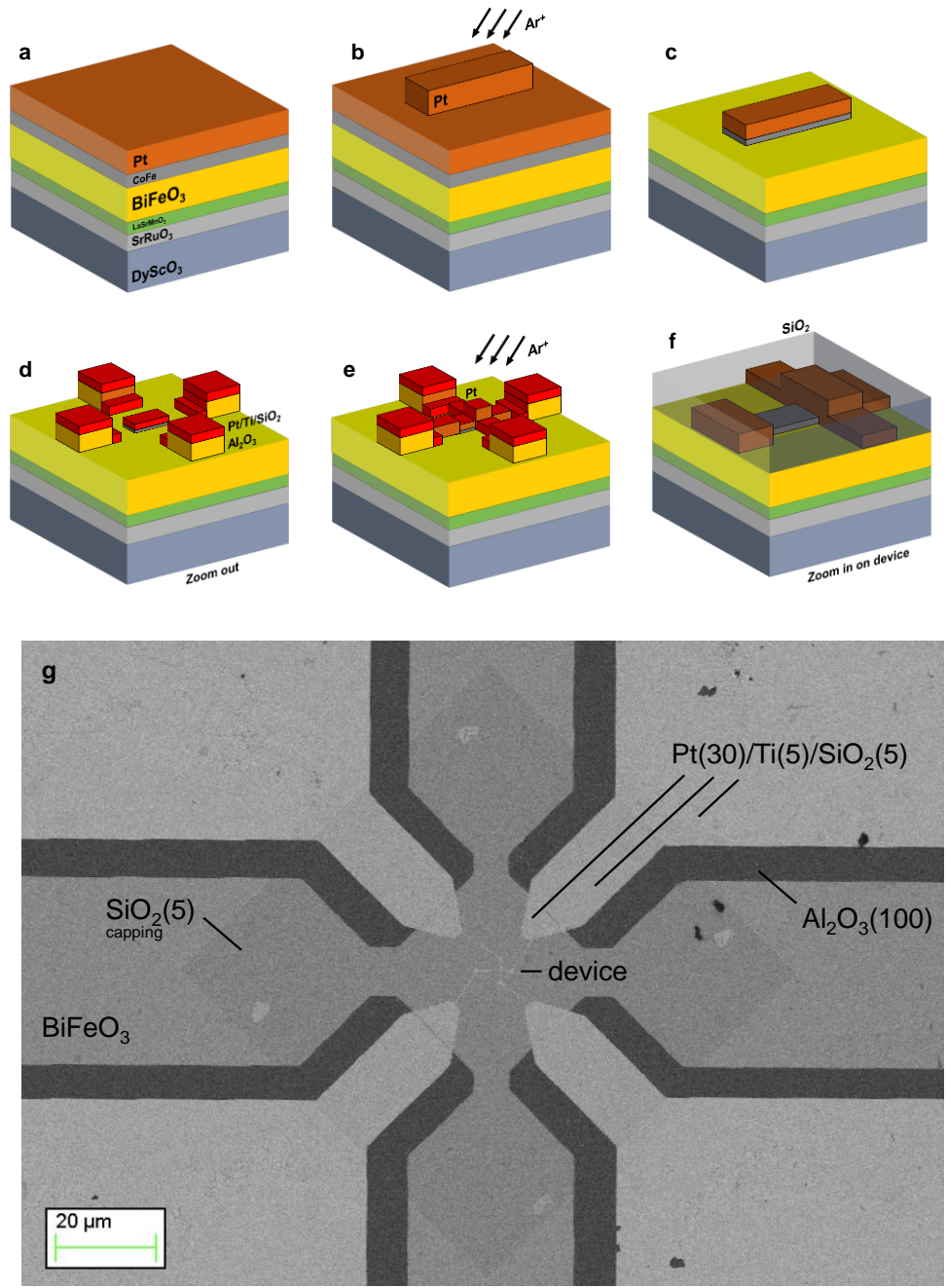

**Figure S1 – Fabrication workflow of MESO nanodevices. a**, Initial full stack. **b**, Patterning of sacrificial Pt nanowire and Ar-ion milling. **c**, Resulting Pt/CoFe wire. **d**, Patterning of contact pads with $Al_2O_3$ and Pt/Ti/$SiO_2$. **e**, Patterning Pt T-shape on the Pt/CoFe wire. **f**, Ar-ion milling of the whole sample and capping with $SiO_2$. **g**, SEM image of the final device.

We start the fabrication of the MESO nanodevices from a full stack of Pt(10 nm)/CoFe(2.5 nm)/BiFeO$_3$(30 nm)/La$_{0.7}$Sr$_{0.3}$MnO$_3$(4 nm)/SrRuO$_3$(10 nm)//DyScO$_3$ (Fig. S1a), deposited and grown with a combination of magnetron sputtering (for the deposition of Pt and CoFe) and pulsed laser deposition (for BiFeO$_3$, La$_{0.7}$Sr$_{0.3}$MnO$_3$, and SrRuO$_3$). The first step of fabrication is the positive e-beam lithography patterning of a nanowire of Pt using PMMA 950A2, onto the initial full stack, which will act as a "sacrificial wire" (Fig. S1b and S1c). The thickness of this wire is 15 nm. Then, Ar-ion milling is used to etch the whole sample, where the time it takes to remove the initial Pt/CoFe layer is the same as for the removal of the fabricated Pt wire. We have opted for this "sacrificial" layer procedure in order to avoid issues of negative resist removal, recurrently observed in lithography processes below 200nm. Then, two positive e-beam lithography steps using double-layer PMMA (495 A4 + 950 A2) are performed to pattern the buffer layer for the bonding pads, made of Al$_2$O$_3$(100 nm), and the contacting pads, made of Pt(40 nm)/Ti(5 nm)/SiO$_2$(5 nm), as seen in Fig. S1d. The use of Al$_2$O$_3$ was crucial in avoiding shortcuts between wire-bonded contacts and the SrRuO$_3$ bottom electrode. Pt/Ti is used to ensure good adhesion between the wire-bonded contact and the pad. SiO$_2$ is used to avoid electrical shortcuts in the areas where Pt/Ti and BiFeO$_3$ are in direct contact. A fourth positive e-beam lithography step is performed to define Pt(20 nm) nanowires, including the T-shaped nanostructure, which are aligned to the initial Pt/CoFe, creating a Pt-on-Pt junction that preserves the original in-situ interface between Pt and CoFe (Fig. S1e). Lastly, the whole sample is etched with Ar-ion milling, so that the Pt on top of the CoFe is removed. The samples are immediately capped with SiO$_2$(5 nm) to prevent CoFe oxidation (Fig. S1f). An SEM image of the full device (including contacts) is shown in Fig. S1g.

**Supplementary Note 2: Extended PFM and MFM characterization.**

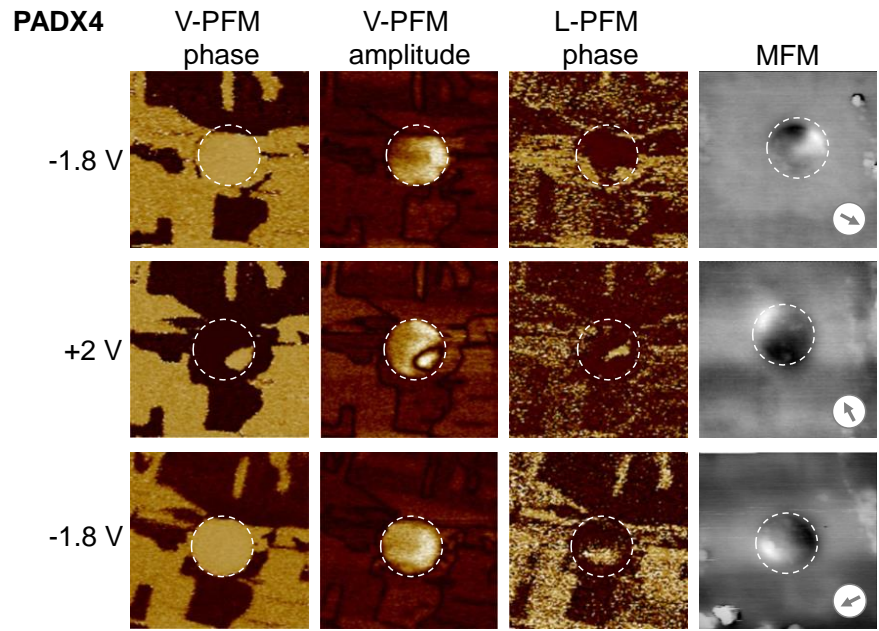

**Figure S2 – Extended PFM and MFM data of Fig. 2 (main text).** Out-of-plane (V) PFM phase and amplitude images and in-plane (L) PFM phase images after applying voltage pulses of −1.8 V, 2 V, and −1.8 V to a disk of Pt/CoFe with a diameter of 200 nm. Corresponding MFM images showing the magnetization direction of the CoFe after each pulse, represented by the grey arrows.

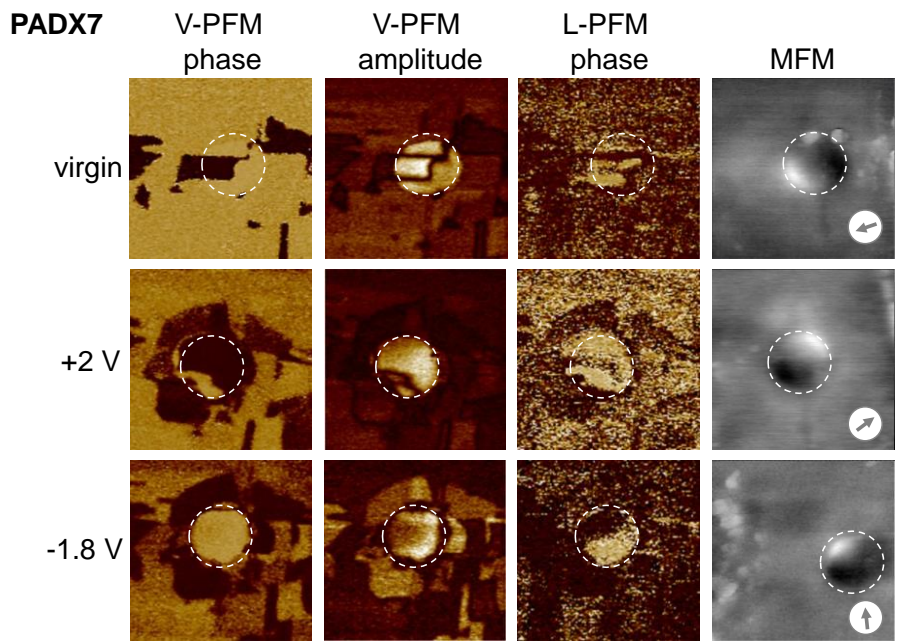

**Figure S3 – Extended PFM and MFM characterization of device X7.** Out-of-plane (V) PFM phase and amplitude images and in-plane (L) PFM phase images for a virgin state and after applying voltage pulses of 2 V and −1.8 V to a disk of Pt/CoFe with a diameter of 200 nm. Corresponding MFM images showing the magnetization direction of the CoFe after each pulse, represented by the grey arrows.

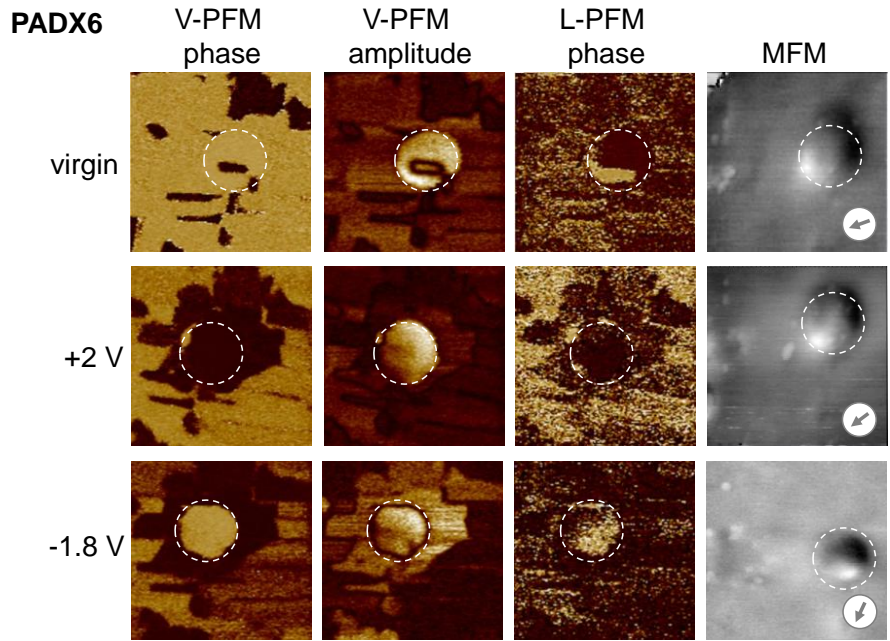

**Figure S4 – Extended PFM and MFM characterization of device X6.** Out-of-plane (V) PFM phase and amplitude images and in-plane (L) PFM phase images for a virgin state and after applying voltage pulses of 2 V and −1.8 V to a disk of Pt/CoFe with a diameter of 200 nm. Corresponding MFM images showing the magnetization direction of the CoFe after each pulse, represented by the grey arrows.

As stated in the main text, out of 9 devices probed (in the same sample) using PFM and MFM, we have observed that the magnetization could be switched only at times and for some devices. In Fig. S2, we show the extended data set for the device discussed in the main text, where magnetization reversal and partial rotation are observed after $V_p$=2 V and $V_p$=−1.8 V, respectively. Device X7, displayed in Fig. S3, showed similar behavior, although the same pulse polarity led to magnetization reversed from the bottom left orientation to the top right. For a subsequent $V_p$=−1.8 V pulse, the magnetization once again rotates, but does not go through a full reversal. We note here that the out-of-plane component of the polarization switches similarly between up and down, while the in-plane component remains rather unchanged, even though with a different contrast than the one shown in Fig. S2 (L-PFM). A third set of data is shown in Fig. S4, for device X6, as an example of a device where no magnetization reversal or rotation was observed, even though the polarization (both out-of-plane and in-plane) was observed to switch. These results emphasize the fact that the manipulation of the magnetization orientation cannot be only connected with the polarization state, and that full control of the magnetization direction will require additional information regarding the magnetic textures of the $BiFeO_3$ underneath the disks.

## Supplementary Note 3: Initialization of the MESO devices

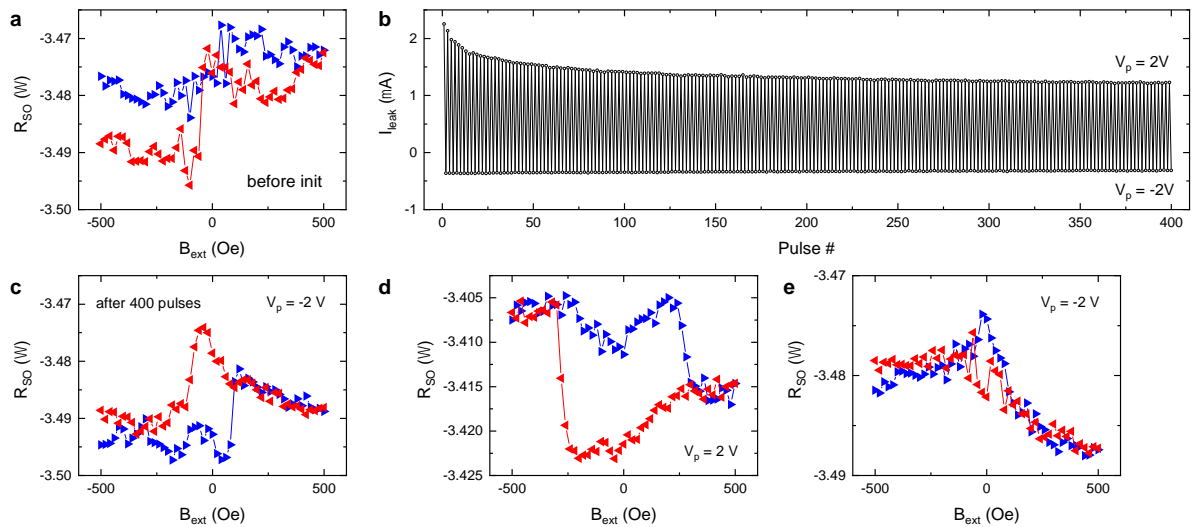

**Figure S5 – Initialization of devices. a**, $R_{SO}$ vs. external magnetic field $B_{ext}$ in the virgin device. **b**, Leakage current $I_{leak}$ for alternating voltage pulses of $V_p=\pm 2$ V. **c**, $R_{SO}$ vs $B_{ext}$ after initialization (last pulse $V_p=-2$ V). **d**, The following $R_{SO}$ vs $B_{ext}$ after $V_p=2$ V. **e**, The following $R_{SO}$ vs $B_{ext}$ after $V_p=-2$ V.

We have consistently observed that the magnetization loops and overall leakage of $BiFeO_3$ change after performing an initialization protocol, where consecutive voltage pulses with opposite polarity are applied. This is consistent with the well-known "wake up" phenomenon in ferroelectrics, related to the depinning of defect-pinned domain walls and removal of charged defects[1]. As shown in Fig. S5a, before initialization, the magnetization loop reveals that the CoFe wire may have pinned magnetic domain regions to the as-grown magnetic state of the $BiFeO_3$. After 400 pulses of $V_p=\pm 2$ V, the leakage current $I_{leak}$ for positive voltage pulses steadily decreases from 2 μA to around 1.25 μA (Fig. S5b). After this, the three possible states discussed in the main text can be observed (Fig. S5c, S5d, and S5e).

# Supplementary Note 4: Correlation between electrical read-out and magnetization orientation.

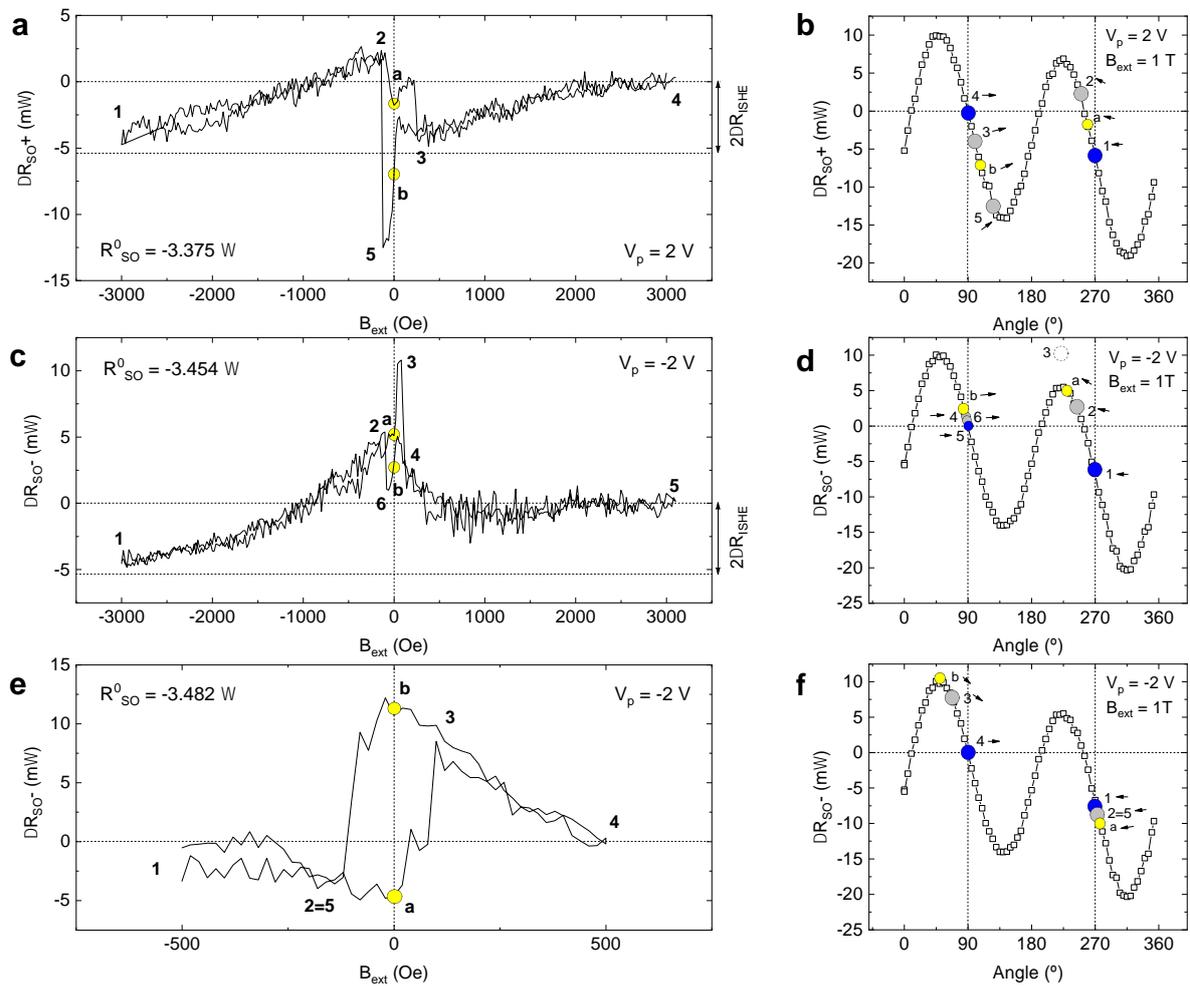

**Figure S6 – Extended magnetization loops and corresponding $R_{SO}$ angle dependence.** $\Delta R_{SO}^{+(-)} = R_{SO} - R^0_{SO}$ vs. external magnetic field $B_{ext}$ and corresponding $\Delta R_{SO}$ angle dependences with $B_{ext} = 1$ T after **a, b,** $V_p = 2$ V, **c, d,** $V_p = -2$ V (intermediate state), **e, f,** and $V_p = -2$ V, where $R_{SO}$ is the raw output resistance, and $R^0_{SO}$ the baseline resistance at large positive $B_{ext}$. The subscript + and – represent the $\Delta R_{SO}$ after $V_p = 2$ V and $V_p = -2$ V, respectively. $\Delta R_{SO}$ values from each angle dependence are numbered and matched to the magnetization loops. Arrows represent the magnetization direction of the CoFe wire as seen from the top.

To obtain further information on which magnetization direction each resistance $R_{SO}$ corresponds to, we have measured the angle dependence of $\Delta R_{SO}$ under a rotating in-plane magnetic field $B_{ext} = 1$ T for $V_p = \pm 2$ V. From $\Delta R_{SO}$ vs. $B_{ext}$, we can identify that $\Delta R_{SO}$ at 3000 Oe corresponds to the magnetization pointing to the right (90°, aligned with the CoFe wire long axis). From this point, we observe if $\Delta R_{SO}$ either increases or decreases when reducing the $B_{ext}$ and match this increase/decrease with the angle dependence, which gives us information on the evolution of the magnetization direction.

# Supplementary Note 5: Full data set for the magnetic switching after $V_p$=2 V.

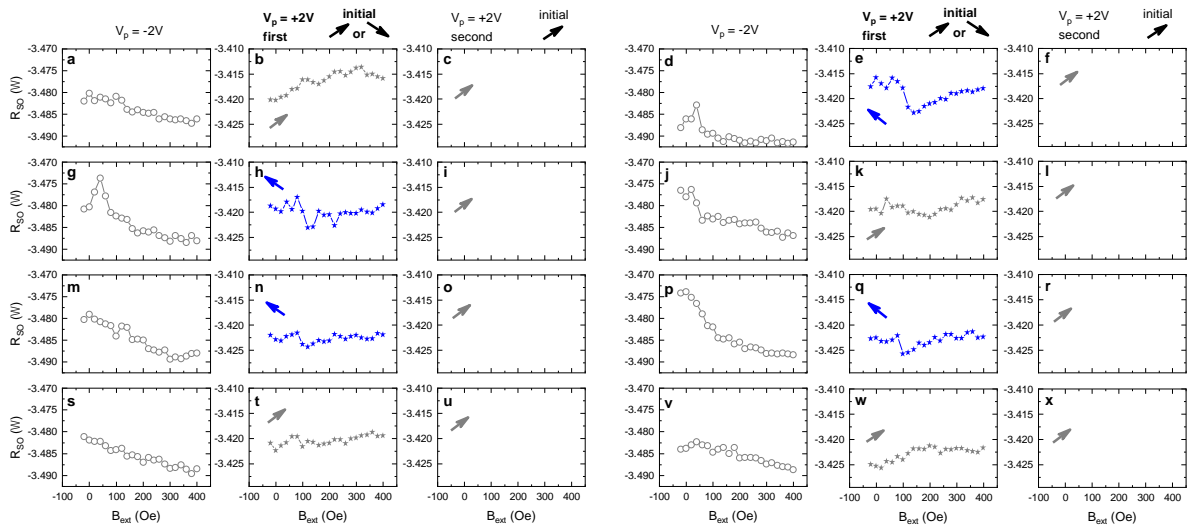

**Figure S7 – Full electrical characterization of the magnetization switching for $V_p$=2 V.** Panels **a** through **x** are shown in the experimental sequential order.

The magnetization is first initialized by applying $V_p$=-2 V (open circles), and the field is swept from 0 Oe to 400 Oe and back. This way, the orientation of the magnetization is known to be oriented towards the right. However, given the two possible states after $V_p$=-2 V, the initial magnetization orientation either points to the top right or bottom right (as shown by the bold arrows at the top). After a first $V_p$=2 V at $B_{ext}$=0 Oe, the magnetization is either observed to switch ~90º (blue stars) or not (grey stars), corresponding to the two $R_{SO}$ branches of Fig. 3d. After collecting this curve, the field is swept back from 400 Oe to 0 Oe, so that the magnetization state now points to the top right. A subsequent $V_p$=2 V never drives additional magnetization rotation (grey full circles).

## Supplementary Note 6: Full data set for the magnetic switching after V$_p$=−2 V.

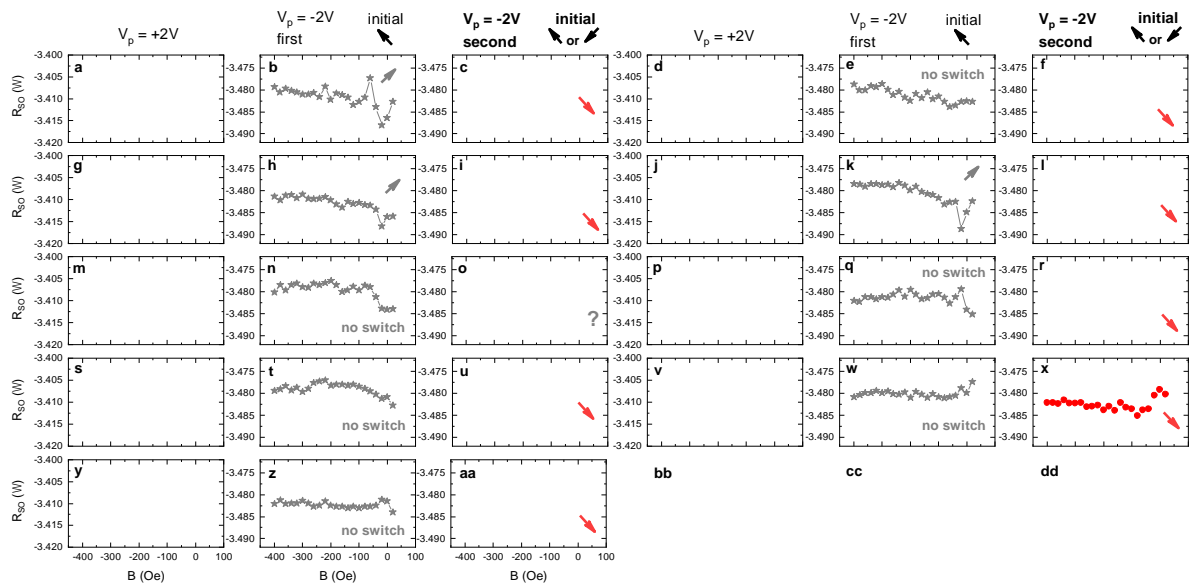

**Figure S8 – Full electrical characterization of the magnetization switching for V$_p$=−2 V.** Panels **a** through **aa** are shown in the experimental sequential order.

The magnetization is first initialized by applying V$_p$=2 V (open circles), and the field is swept from 0 Oe to −400 Oe and back. This way, the orientation of the magnetization is oriented towards the top left. Then, after a first V$_p$=−2 V at B$_{ext}$=0 Oe (grey stars), the magnetization is at times observed to switch by ~90º (Fig. S8b, S8h, S8k), given that the sweep to −400 Oe reveals a small jump in R$_{SO}$ as seen in Fig. 3e. The remaining curves (labeled "no switch") may correspond to a slight rotation of the magnetization to the bottom left, as seen in Fig. 3f. After collecting this curve, the field is swept back from −400 Oe to 0 Oe, so that the magnetization state now points to either the top left or bottom left (as shown by the bold arrows at the top), depending on whether V$_p$=−2 V led to the intermediate of fully switched state (Fig. 3e or 3f, respectively). A subsequent V$_p$=−2 V now drives a magnetization reversal towards the right (red full circles), corresponding to the upper branch in Fig. 3f. This behavior is always observed, except for one instance (Fig. S8o) where it is unclear if the magnetization is fully reversed.

**Supplementary Note 7: Full data set of the read-out resistance after $V_p=\pm2$, with and without a static magnetic field applied.**

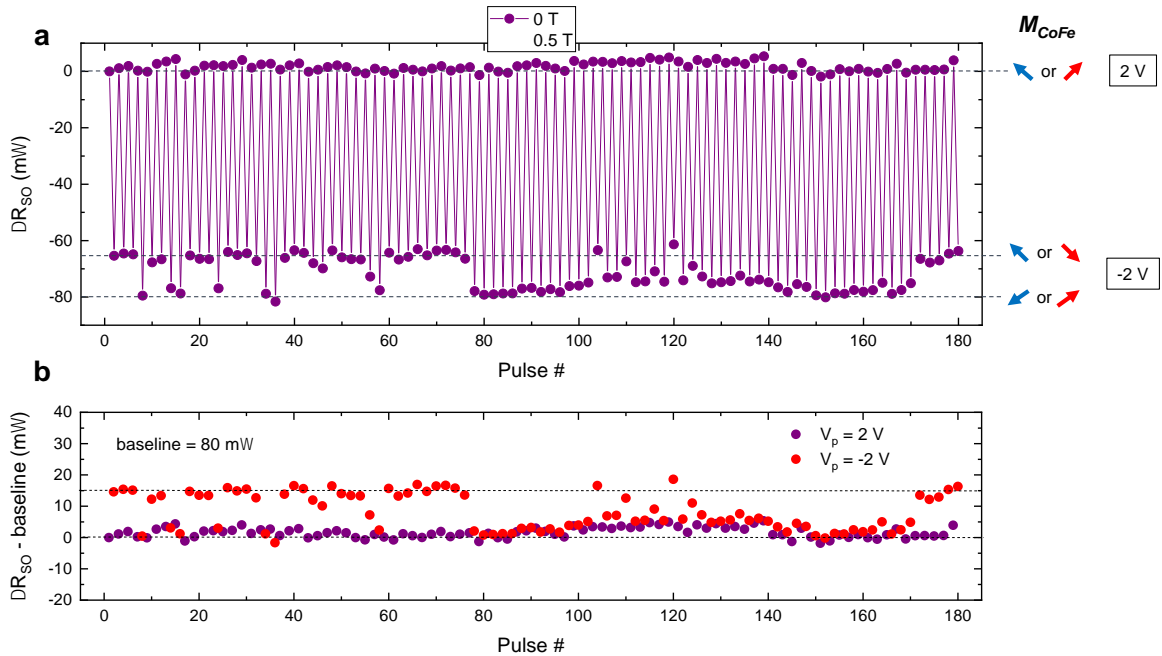

**Figure S9 – Steady-state switching of the MESO device. a,** $\Delta R_{SO}$ as a function of alternating $V_p=\pm2$ V at 0 T (purple filled circles) and 0.5 T (empty circles). Blue and red arrows on the right represent the possible CoFe magnetization orientations as seen from the top of the device. **b**, Same data as presented in panel a, but subtracted by the baseline contribution of about 80 mΩ.

Using the MESO device as an active element in circuits will require the observation of different output states for different magnetization direction. In Fig. S9a, we plot the output resistance $\Delta R_{SO}$, normalized by the average output resistance after $V_p=2$ V. In the main text, although we have used a magnetic field sweep to verify the magnetization direction upon voltage pulse, switching and reading functionality of the MESO device should ideally be shown without any field applied, since that is how the device should work in a real circuit scenario.

We observe that, between voltage pulses of opposite polarity, $\Delta R_{SO}$ shifts around −80 mΩ, corresponding to the baseline shift discussed in the main text and shown in Fig. 3a. We reemphasize that the baseline shift is attributed to a slight modulation of the resistivity of CoFe and not magnetization switching.

However, on top of this shift, we observe that after $V_p=-2$ V two resistances values ($\Delta R_{SO}=-65$ mΩ and $\Delta R_{SO}=-80$ mΩ) are obtained. Without the external magnetic field sweep analysis, the two possible states observed in Fig. 4 (main text) make it difficult to identify to which magnetization direction each resistance value corresponds. As shown by the arrows next to the plot, $\Delta R_{SO}=-65$ mΩ can represent magnetization pointing towards the top left or bottom right, while $\Delta R_{SO}=-80$ m$\Omega$ can represent magnetization pointing towards the bottom left or top right.

For $V_p$=2 V, where only one state is observed, the variability of $\Delta R_{SO}$ after each pulse is in the same order of magnitude as the expected $R_{SO}$ difference between the two possible magnetization states (magnetization towards the top left or top right, as represented by the arrows).

Lastly, by performing the same experiment under an in-plane $B_{ext}$=0.5 T (open circles), i.e. fixing the magnetization direction in one direction throughout the whole experiment, $\Delta R_{SO}=-65$ m$\Omega$ is never observed, suggesting that while the $BiFeO_3$ is still switching (leading to a baseline shift of ~80 m$\Omega$), the extra change in $\Delta R_{SO}$ may be related with changes in the magnetization direction of the CoFe.

The change in $\Delta R_{SO}$ is further evidenced in Fig. S9b, where the baseline is subtracted to the data presented in panel a. Here, we observe that, at times, a shift of about 15 m$\Omega$ is observed upon application of a $V_p=-2$ V. However, given the uncertainty associated with different magnetization orientations corresponding to similar $\Delta R_{SO}$, a magnetic field sweep is still required to ensure that magnetization switching occurred, as presented in Fig. 4 of the main text. As a counterexample, solely based on the $\Delta R_{SO}$ data presented here, we may have a tilt of the magnetization of <45º, which could represent a shift in $\Delta R_{SO}$ of about 15 m$\Omega$ (see Fig. 3c, in the main text).

**Supplementary Note 8: Effect of voltage pulse duration on the switching of the BiFeO$_3$ and the switching of M$_{CoFe}$.**

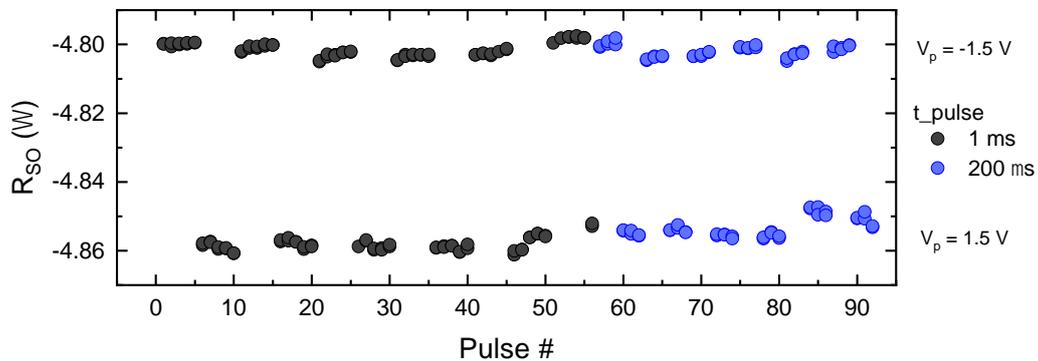

**Figure S10** – Output resistance R$_{SO}$ as a function of alternating V$_p$=±2 V at 0 T, for a pulse duration time of 1 ms (dark circles) and 200 μs.

All the switching experiments shown in the main text were performed with a voltage pulse duration of 200 μs, mostly due to the time constraints of running the long full set of experiments for several devices at different pulse durations. As displayed in Figure S10, we observed no change in the ability to switch the BiFeO$_3$ when increasing the voltage pulse duration from 200 μs to 1 ms. However, to minimize excessive current flow through the BiFeO$_3$, and to avoid the potential breakdown of this layer, we have used the minimum pulse duration experimentally possible by the equipment used.

However, the discussion on the effect of voltage pulse duration on these devices can be divided in two parts: 1) the switching dynamics of the BiFeO$_3$, and 2) the switching dynamics of the CoFe magnetic element upon switching the BiFeO$_3$.

1) It was experimentally demonstrated[2] that a 20-nm-thick BiFeO$_3$ can be switched with less than 10 ns voltage pulses, with pulse duration being linearly dependent on the lateral area of the top electrode. A rough comparison with our Pt/CoFe-based MESO devices reveals that with an approximate area of 190 μm$^2$ (accounting for the Pt nanowires that are also in direct contact with the BiFeO$_3$), our 30-nm-thick BiFeO$_3$ could, in principle, be switched with a pulse as short as ~50 ns.

2) Yet, it was theoretically shown that when considering a CoFe/BiFeO$_3$ system, the switching time to successfully switch the CoFe magnetic element also needs to be considered[3]. While the lowest theoretical switching time for BiFeO$_3$ was 30 ps, the minimum time required to switch the magnetic element was 1.45 ns, meaning that the multiferroic cannot switch too fast, or the M$_{CoFe}$ will not be able to follow it. We note that these simulations were performed for a magnetic element of 40 nm x 20 nm (thickness of 2 nm), while our CoFe element is 500 nm x

150 nm (thickness of 2.5 nm), implying that the minimum switching time could be much larger in our case.

With this in mind, we expect that the 200 µs used in the experiments shown in the main text is well above the expected minimum switching time of $M_{CoFe}$, but not high enough to promote larger leakage or potential damage to the $BiFeO_3$ layer.

**Supplementary Note 9: Coupling between different sketched CoFe magnetic elements with BiFeO$_3$.**

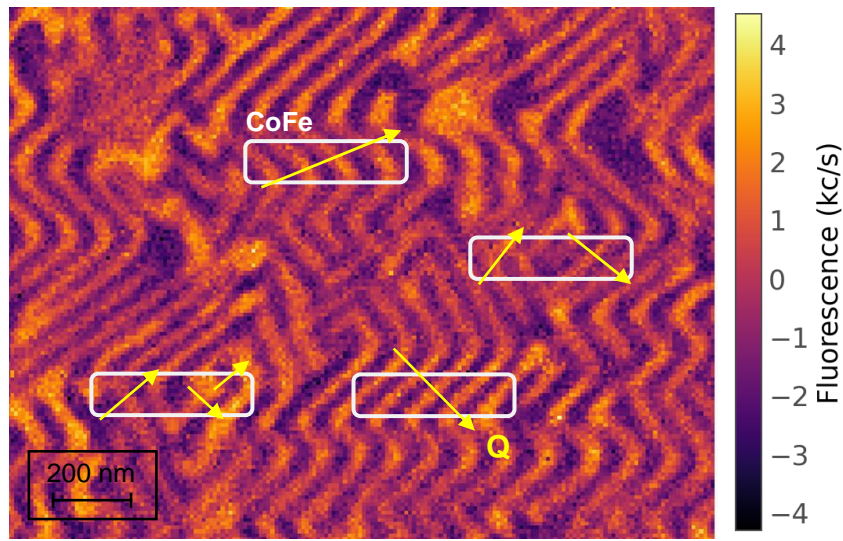

**Figure S11** – Sketch of randomly distributed CoFe elements, to scale, with respect to the magnetic landscape of BiFeO$_3$.

The lack of device-to-device reproducibility can be attributed to the complex magnetic textures and spin cycloid probed by nitrogen-vacancy (NV) magnetometry and displayed in Fig. 5 of the main text. As sketched in Fig. S11, when fabricating the MESO devices, we have no knowledge of the position of the device with respect to this magnetic texture, leading to a situation where several devices with the exact same orientation, but fabricated a few microns apart, will experience completely different exchange coupling with BiFeO$_3$. To improve the reproducibility of these devices, the ferroelectric domain structure of BiFeO$_3$ needs to be engineered to ideally possess a single macroscopic domain region, with a coherent cycloid propagation that can be consistently switched. Alternatively, having no cycloid present, i.e., a uniform magnetic region where M$_{CoFe}$ would couple with a uniform antiferromagnetic order in the multiferroic, could vastly simplify the reproducibility issues. Both approaches have been experimentally explored through strain and chemical doping[4], but have yet to be implemented with functional nanodevices.

**Supplementary references**